\documentclass[mathleft
]{an}
\usepackage{graphicx}
\usepackage{times}
\begin{document}

\Pagespan{1}{}
\Yearpublication{2011}%
\Yearsubmission{2011}%
\Month{9}%
\Volume{332}%
\Issue{9/10}%
 \DOI{10.1002/asna.201111597}%

\title{Chemical composition of AY\,Ceti: A flaring, 
spotted star with a white dwarf companion 
}

\author{G.~Tautvai\v{s}ien\.{e}\inst{1}\fnmsep\thanks{Corresponding author:
  {grazina.tautvaisiene@tfai.vu.lt}}
\and G.~Barisevi\v{c}ius\inst{1}
\and S.~Berdyugina\inst{2}
\and I.~Ilyin\inst{3}
\and Y.~Chorniy\inst{1} 
}
\titlerunning{Chemical composition of AY\,Ceti}
\authorrunning{G. Tautvai\v{s}ien\.{e} et al.}
\institute{
Institute of Theoretical Physics and Astronomy, Vilnius University,
Go\v{s}tauto 12, LT-01108 Vilnius, Lithuania
\and
Kiepenheuer Institut f\"ur Sonnenphysik, Sch\"oneckstr 6, D-79104 Freiburg, 
Germany
\and
Leibniz-Institut f\"ur Astrophysik Potsdam, An der Sternwarte 16, 
D-14482 Potsdam, Germany
}

\received{2011 Sep 20}
\accepted{2011 Sep 29}
\publonline{2011}


\abstract{%
The detailed chemical composition of the atmosphere AY\,Cet (HD\,7672) is determined 
from a high-resolution spectrum in the optical region. The main atmospheric 
parameters and the abundances of 22 chemical elements, including key species 
such as $^{12}{\rm C}$, $^{13}{\rm C}$, N, and O, are determined.
A differential line analysis gives $T_{\rm eff}=5080$~K, log\ $g=3.0$, ${\rm
[Fe/H]}=-0.33$, ${\rm [C/Fe]}=-0.17$, ${\rm [N/Fe]}=0.17$, ${\rm
[O/Fe]}=0.05$, ${\rm C/N}=1.58$, and $\rm \textsuperscript{12}C/\textsuperscript{13}C=21$. 
Despite the high chromospheric activity,  the optical spectrum of AY\,Cet provides 
 a chemical composition typical for first ascent giants after 
the first dredge-up. 
}

\maketitle

\section{Introduction}
The variability of
AY\,Ceti (HD\,7672, 39\,Cet) was first noticed by Cousins (1962) and later confirmed 
by Olsen (1974).  A spectral class of G5\,III was assigned to AY\,Cet by Cowley \& Bidelman (1979), who 
also noted in its spectrum a moderately strong  Ca\,{\sc ii} H and K emission, indicating that  this star 
might be an RS CVn binary.  

From photoelectric photometry obtained in 1971, 1972, 1980--81, and 1981--82, Eaton et al.\ (1983) have 
determined a mean photometric period of ${77.65 \pm 0.05}$~days. The full $V$ amplitude of the nearly 
sinusoidal light curve has been as large as 0.18~mag. For predicting future behaviour of AY\,Cet the authors 
proposed the ephemeris JD $=244636.0 + 77.65\,E$ days, where the initial epoch is at time of minimum light. 
In the same work it was noticed that the maximum brightness of AY\,Cet 
has decreased by 0.18~mag (from  $V=5.35$ to 5.53~mag) during the ten years of observations. 
Strassmeyer et al.\ (1989) found a slightly smaller  
rotational period of $75.12\pm 0.03$~days from photometry obtained between 1983 and 1986. The amplitude 
was variable from season to season and changed from approximately 0.06 to 0.19~mag within one year.  
Poretti et al.\ (1986) have combined $U\!BV$ photometry of AY\,Cet from 1980--81, 1981--82, 
and 1983--84, and  found that the variations of the $V$ magnitude can be represented as a combination of three 
sine curves with the following semi-amplitudes and periods: 0.06~mag at ${P_1=77.22}$~d, 0.03~mag at 
$P_2=79.36$~d, and 0.07~mag at $P_3=1820$~d. 

As well as for $\lambda$\,And, the orbital period of AY\,Cet is much smaller than the photometric period. 
Using 22 measurements, Simon et al.\ (1985) have determined for AY\,Cet an 
orbital period of $56.80\pm 0.03$ days, which attributes AY\,Cet to the sample of non-synchronous 
RS CVn rotators.  Even more accurate orbital elements have been determined for AY\,Cet by 
Fekel \& Eitter (1989). 
    
From observations with the HEAO\,II satellite (Walter \& Bowyer 1981) and from IUE observations 
(Simon et al.\ 1985) it was discovered that AY\,Cet has a hot white dwarf companion. The UV spectrum of 
AY\,Cet at 1200--2000~\AA\  shows a hot stellar continuum and a very broad Ly$\alpha$ absorption line which 
can be matched to the energy distribution of the white dwarf with ${T_{\rm eff}=18\,000}$~K and 
$\log g=8$ model 
atmosphere.  Superimposed on the hot continuum are high-excitation emission lines of the late-type primary 
(e.g., O\,{\sc i} 1305~\AA\ and C\,{\sc iv} 1549~\AA). Simon et al.\ determined that the radius of the AY\,Cet\,B white 
dwarf  is $r_2=0.012\ \rm R_{\odot}$, and the mass ${m_2=0.55}\ \rm M_{\odot}$, the radius of the primary AY\,Cet\,A is  
$r_1=6.8\ \rm R_{\odot}$, and the mass $m_1=2.09\ \rm M_{\odot}$. Mass and radius of AY\,Cet were also determined in 
several later studies: $r_1=6.6\ \rm R_{\odot}$  by Schrijver \& Zwaan (1991); $r_1=8.18 \ \rm R_{\odot}$ 
and  $m_1=1.95\ \rm M_{\odot}$ by da~Silva et al.\ (2006); and $r_1=7.9\ \rm R_{\odot}$ and  $m_1=2.0 \ \rm M_{\odot}$  
 by Massarotti et al.\ (2008).   
 
From observations with the VLA radio interferometer, Simon et al.\ (1985) also reported the detection of two 
very strong radio flares on AY\,Cet. The second flare was characterized by minute-to-minute variations in the 
left-circularly polarized flux.  Such highly polarized activity is quite rare among the RS CVn and related systems 
and most probably acts as electron-cyclotron maser. In order to study weather the UV emission of AY\,Cet is 
variable, Simon \& Sonneborn (1987) have obtained IUE observations of this star on five dates between 1983 
and 1984. It was found that the spectrum of the secondary has no discernible changes either in the level of 
brightness of its continuum or in the shape of the absorption profile of Ly$\alpha$. Quite significant changes 
were observed in the spectrum of the primary. At their strongest, the stellar features were up to one hundred 
times brighter than the same lines  observed in the quiet Sun. The range in line strength from the minimum to 
the maximum brightness levels observed for AY\,Cet was roughly a factor of 1.5--2 in the chromospheric lines and 
5--6 in the transition region features. Surprisingly, both UV and previously observed radio flares 
(Simon et al.\ 1985) occurred near the peak in visible light curve, when a surface of the primary is less 
covered by starspots.        

Thus, AY\,Ceti shares the attributes of spotted, chromospherically active RS CVn binaries: strong 
Ca\,{\sc ii} H and K emission, a prominent photometric wave in visual light, and intense soft X-ray emission. 
In addition, AY\,Cet  is especially prominent because of its strong radio flares. From the radio point of view, 
AY\,Cet may be among the most active stellar sources in the sky (Simon et al.\ 1985).  The purpose of this work 
was to determine the detailed chemical composition of AY\,Cet atmosphere from its high-resolution spectrum in the 
optical region.  In particular, it was interesting to investigate has stellar activity altered abundances of 
mixing-sensitive elements such as  $^{12}{\rm C}$, $^{13}{\rm C}$,  and N.  

\section{Observations and method of analysis}
A spectrum of AY\,Cet was observed in August of 1999  
on the 2.56~m Nordic Optical Telescope using the SOFIN echelle spectrograph with 
the optical camera, which provided a spectral resolving power of ${R \approx
80\,000}$, for 26 slightly shifted in wavelength spectral orders, each of $\sim$40~\AA, in the spectral
region from 5000 to 8300~\AA.  The details of spectral reductions were
presented in Tautvai\v{s}ien\.{e} et al.\ (2010, hereafter Paper~I).
%
We selected 147 atomic lines for the measurement of
equivalent widths and 16 lines for the comparison with
synthetic spectra. The measured equivalent widths of lines are presented in Table~1. 

The spectra were analysed using a differential model
atmosphere technique described in Paper~I.  Here we present
only some details.

\begin{table*}
\caption{Measured equivalent widths of lines, $EW$, in the AY\,Cet spectrum.}
\label{Table1}
\begin{tabular}{ccc@{\hspace{15mm}}ccc@{\hspace{15mm}}ccc}
\hline\noalign{\smallskip}
Element & $\lambda$\,(\AA) & $EW$\,(m\AA) & Element & $\lambda$\,(\AA) &$EW$\,(m\AA) &Element &$\lambda$\,(\AA) &$EW$\,(m\AA)\\[1.5pt]
\hline\noalign{\smallskip}
~~Si\,{\sc i}	 &    5517.55~~ &   15.2 & 	 &    6266.30~~ &   16.8 & 	 &    6646.97~~ &   21.4 \\
	 &    5645.60~~ &   38.2 & 	 &    6274.66~~ &   33.0 & 	 &    6786.86~~ &   30.5 \\
	 &    5665.55~~ &   38.7 & 	 &    6285.16~~ &   42.6 & 	 &    6793.27~~ &   15.1 \\
	 &    5793.08~~ &   42.8 & 	 &    6292.82~~ &   50.6 & 	 &    6839.83~~ &   53.7 \\
	 &    5948.54~~ &   77.9 & ~~Cr\,{\sc i}	 &    5712.78~~ &   25.9 & 	 &    6842.69~~ &   44.2 \\
	 &    6131.85~~ &   20.6 & 	 &    5783.87~~ &   54.5 & 	 &    6843.65~~ &   65.6 \\
	 &    7003.57~~ &   45.5 & 	 &    5784.97~~ &   44.9 & 	 &    6851.64~~ &   14.0 \\
~~Ca\,{\sc i}	 &    5260.38~~ &   44.2 & 	 &    5787.92~~ &   56.0 & 	 &    6857.25~~ &   27.9 \\
	 &    5867.57~~ &   32.0 & 	 &    6661.08~~ &   14.1 & 	 &    6858.15~~ &   54.4 \\
	 &    6455.60~~ &   73.8 & 	 &    6979.80~~ &   48.2 & 	 &    6862.49~~ &   31.9 \\
	 &    6798.47~~ &   10.5 & 	 &    6980.91~~ &   12.7 & 	 &    7461.53~~ &   48.3 \\
~~Sc\,{\sc ii}	 &    5526.81~~ &  103.7 & ~~Fe\,{\sc i}	 &    5395.22~~ &   25.0 & ~~Fe\,{\sc ii}	 &    5132.68~~ &   35.1 \\
	 &    5640.98~~ &   65.6 & 	 &    5406.78~~ &   44.0 & 	 &    5264.81~~ &   50.0 \\
	 &    5667.14~~ &   52.0 & 	 &    5522.45~~ &   51.3 & 	 &    6113.33~~ &   13.0 \\
	 &    6279.75~~ &   51.7 & 	 &    5577.03~~ &   11.1 & 	 &    6369.46~~ &   22.1 \\
	 &    6300.69~~ &   11.2 & 	 &    5579.35~~ &   13.7 & 	 &    6456.39~~ &   67.9 \\
~~Ti\,{\sc i}	 &    5648.58~~ &   22.1 & 	 &    5607.67~~ &   19.1 & 	 &    7711.72~~ &   48.1 \\
	 &    5662.16~~ &   43.9 & 	 &    5608.98~~ &   14.0 & ~~Co\,{\sc i}	 &    5530.78~~ &   36.3 \\
	 &    5716.45~~ &   14.5 & 	 &    5651.48~~ &   23.3 & 	 &    5590.71~~ &   27.0 \\
	 &    5739.48~~ &   18.6 & 	 &    5652.33~~ &   34.0 & 	 &    5647.23~~ &   28.3 \\
	 &    5880.27~~ &   28.5 & 	 &    5653.86~~ &   45.2 & 	 &    6117.00~~ &   16.3 \\
	 &    5899.30~~ &   70.5 & 	 &    5679.03~~ &   66.8 & 	 &    6188.98~~ &   25.2 \\
	 &    5903.31~~ &   18.6 & 	 &    5720.90~~ &   19.0 & 	 &    6455.00~~ &   19.7 \\
	 &    5941.76~~ &   48.8 & 	 &    5732.30~~ &   12.8 & 	 &    6595.86~~ &    8.6 \\
	 &    5953.17~~ &   60.0 & 	 &    5741.86~~ &   36.9 & 	 &    6678.82~~ &   13.7 \\
	 &    5965.83~~ &   57.3 & 	 &    5784.67~~ &   39.8 & ~~Ni\,{\sc i}	 &    5578.73~~ &   77.3 \\
	 &    6064.63~~ &   30.1 & 	 &    5793.92~~ &   40.9 & 	 &    5587.87~~ &   77.6 \\
	 &    6098.66~~ &   10.5 & 	 &    5806.73~~ &   62.3 & 	 &    5589.37~~ &   27.2 \\
	 &    6121.00~~ &   11.9 & 	 &    5807.79~~ &   15.7 & 	 &    5593.75~~ &   41.4 \\
	 &    6126.22~~ &   54.2 & 	 &    5809.22~~ &   63.7 & 	 &    5643.09~~ &   13.6 \\
	 &    6220.49~~ &   20.8 & 	 &    5811.92~~ &   15.4 & 	 &    5748.35~~ &   50.7 \\
	 &    6303.77~~ &   25.1 & 	 &    5814.82~~ &   30.6 & 	 &    5805.22~~ &   38.6 \\
	 &    6599.11~~ &   31.7 & 	 &    6027.06~~ &   74.5 & 	 &    6053.68~~ &   19.1 \\
	 &    6861.45~~ &   17.3 & 	 &    6034.04~~ &   10.9 & 	 &    6108.12~~ &   88.7 \\
~~V\,{\sc i}	 &    5604.96~~ &   14.1 & 	 &    6035.35~~ &    7.8 & 	 &    6111.08~~ &   34.7 \\
	 &    5646.11~~ &   16.4 & 	 &    6054.07~~ &   11.4 & 	 &    6128.98~~ &   44.3 \\
	 &    5657.45~~ &   22.5 & 	 &    6056.01~~ &   73.2 & 	 &    6130.14~~ &   19.8 \\
	 &    5668.37~~ &   20.8 & 	 &    6098.25~~ &   20.5 & 	 &    6204.60~~ &   23.5 \\
	 &    5670.86~~ &   49.2 & 	 &    6105.13~~ &   14.1 & 	 &    6378.25~~ &   33.7 \\
	 &    5727.66~~ &   31.1 & 	 &    6120.24~~ &   20.2 & 	 &    6482.80~~ &   59.6 \\
	 &    5737.07~~ &   35.8 & 	 &    6187.99~~ &   58.3 & 	 &    6586.32~~ &   64.9 \\
	 &    5743.43~~ &   26.4 & 	 &    6200.32~~ &   98.6 & 	 &    6598.60~~ &   22.6 \\
	 &    6039.74~~ &   35.6 & 	 &    6226.74~~ &   34.8 & 	 &    6635.13~~ &   20.7 \\
	 &    6058.18~~ &   11.8 & 	 &    6229.23~~ &   56.0 & 	 &    6767.78~~ &  111.1 \\
	 &    6111.65~~ &   36.4 & 	 &    6270.23~~ &   72.4 & 	 &    6772.32~~ &   54.7 \\
	 &    6119.53~~ &   51.6 & 	 &    6380.75~~ &   60.1 & 	 &    6842.03~~ &   32.7 \\
	 &    6135.37~~ &   32.7 & 	 &    6392.54~~ &   33.2 & 	 &    7001.55~~ &   25.1 \\
	 &    6224.50~~ &   27.8 & 	 &    6574.21~~ &   62.8 & 	 &    7062.97~~ &   25.0 \\
	 &    6233.19~~ &   25.9 & 	 &    6581.21~~ &   46.8 & 	 &    7715.59~~ &   55.5 \\
\hline
\end{tabular}
\end{table*}

\subsection{Atmospheric parameters}
Initially, the effective temperature, $T_{\rm eff}$, of AY\,Cet was
derived and averaged from the intrinsic color indices $(B-V)_0$ and
$(b-y)_0$ using the corrected calibrations by Alonso et al.\ (1999).
The  color indices $B-V=0.89$ and $b-y=0.558$ were taken from 
Mermilliod (1986) and Hauck \& Mermilliod (1998), respectively.
A reddening of $E_{B-V}=0.02$, estimated using the Hakkila et
al.\ (1997) software, was taken into account. The temperatures deduced from both 
color indices were 5080~K and,  no obvious trend
of the Fe\,{\sc i} abundances with the excitation potential was found
(Fig.~1).

The surface gravity log\,$g$ was found by adjusting the model's gravity to
yield the same iron abundance from the Fe\,{\sc i} and Fe\,{\sc ii}
lines.  The microturbulent velocity $v_{\rm t}$ value corresponding to a
minimal line-to-line Fe\,{\sc i} abundance scattering was chosen as a
correct value.  Consequently, [Fe/H] values do not depend on the
equivalent widths of lines (Fig.~2).

\begin{figure}
\label{Figure1}
\includegraphics[width=82mm,height=40mm]{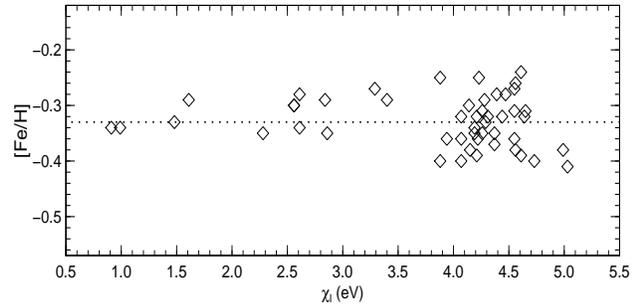}
\caption{
The [Fe\,{\sc i}/H] abundance values versus the lower
excitation potential $\chi_{\rm l}$.  The mean abundance
[Fe\,{\sc i}/H] $=-0.33$ is shown as a dotted line.
}
\end{figure}

\begin{figure}
\label{Figure2}
\includegraphics[width=82mm,height=40mm]{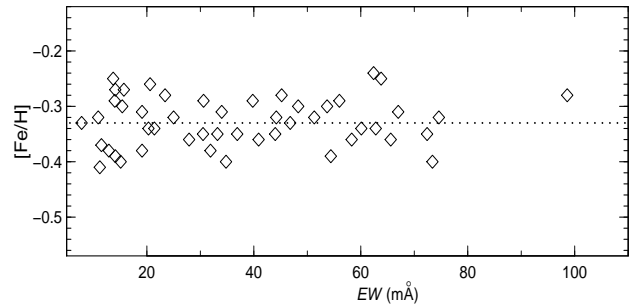}
\caption{
The [Fe\,{\sc i}/H] abundance values versus
the equivalent widths. The mean abundance 
[Fe\,{\sc i}/H] $=-0.33$ is shown as a dotted line.
}
\end{figure}

\subsection{Mass determination}
The mass of AY\,Cet was evaluated from its effective temperature,
luminosity and the isochrones from Girardi et al.\ (2000).  The
luminosity $\log\,(L/L_{\odot})=1.65$ was calculated from the {Hipparcos} 
parallax $\pi=12.41$~mas (van Leeuwen 2007) and $V =
5.43$~mag (da Silva et al.\  2006),
the bolometric correction calculated according to Alonso et al.\
(1999), and the above mentioned reddening ${E_{B-V}=0.02}$.  A mass of 
$\sim$\,2.2\,M$_{\odot}$ was found, which is close to the previous 
evaluations (Simon et al.\ 1985; da Silva et al.\ 2006; Massarotti et al.\ 2008). 

\subsection{Spectrum syntheses}
Due to the rotation spectral lines of AY\,Cet are slightly broadened.  
We used ${v\,{\rm sin}\,i=4.0}~{\rm km\,s}^{-1}$ because with this value the match 
between the synthetic 
and observed spectra was the best. The same rotation speed was determined by 
Strassmeier et al.\ (1988), Berdyugina (1994), and Montes et al.\ (1994). 
 This value is also close to $v\,{\rm sin}\,i=4.5~{\rm km\,s}^{-1}$ determined 
by Gray (1989).  A slightly lower value of $v\,{\rm sin}\,i=2.9~{\rm km\,s}^{-1}$  
was used by Batten et al.\ (1989), and slightly higher value of 
 $v\,{\rm sin}\,i=6.0~{\rm km\,s}^{-1}$ by Fekel et al.\ (1986).

The method of synthetic spectra was used to determine the carbon abundance
from the C$_2$ line at 5135.5~{\AA}.  The interval 7980--8130~{\AA},
containing strong $^{12}$C$^{14}$N and $^{13}$C$^{14}$N features, was
used for the nitrogen abundance and $^{12}$C/$^{13}$C ratio
determinations.  The $^{12}$C/$^{13}$C ratio was determined
from the (2,0) $^{13}$C$^{12}$N feature at 8004.7~{\AA}.  All
log\,$gf$ values were calibrated to fit to the solar spectrum of Kurucz
(2005) with solar abundances from Grevesse \& Sauval (2000).  

\begin{figure*}
\label{Figure3}
\includegraphics[width=135mm]{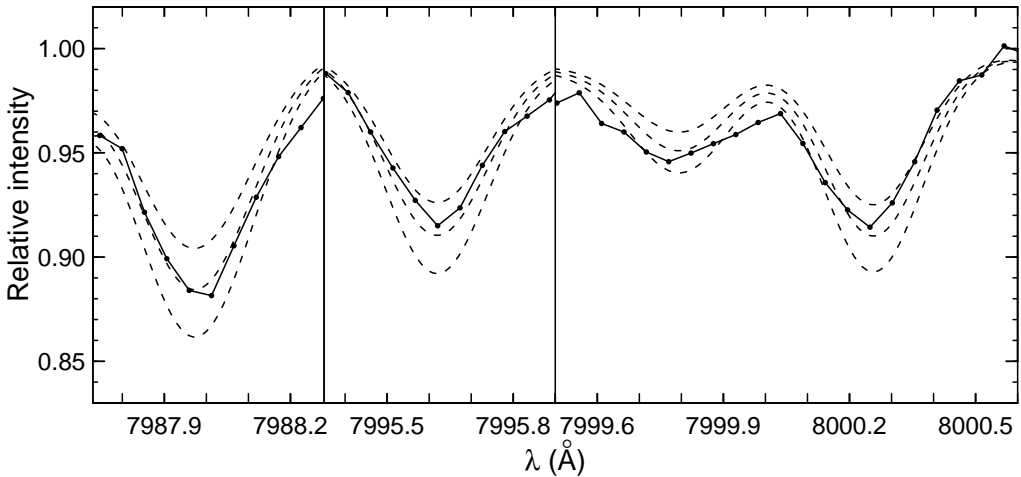}
\caption{
Synthetic spectrum fit to $^{12}$CN lines.
The observed spectrum is shown as a solid line. The dashed lines are
synthetic spectra with ${\rm [N/Fe]} = 0.13$, 0.23, and 0.33~dex.
}
\end{figure*}

The oxygen abundance was determined from the forbidden [O\,{\sc i}] line
at 6300.31~\AA\ with the oscillator strengths  for
$^{58}$Ni and $^{60}$Ni from Johansson et al.  (2003) and 
log~$gf = -9.917$ obtained by fitting to the solar spectrum (Kurucz
2005) with ${\log A_{\odot}=8.83}$ (Grevesse \& Sauval 2000).
In Figs.~3, 4, and 5 we show several examples of synthetic spectra in the 
vicinity of the $^{12}$C$^{14}$N, C$_2$, and [O{\sc i}] lines.

\begin{figure}
\label{Figure4}
\includegraphics[width=70mm]{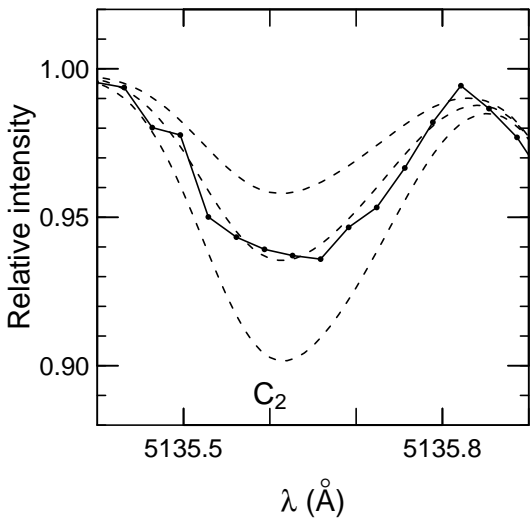}
\caption{
Synthetic spectrum fit to the C$_2$ line at 5135.5~{\AA}. The observed spectra
are shown as solid lines. The  dashed lines are synthetic spectra with
${\rm [C/Fe]} =-0.03$, 0.13, and 0.23~dex.
}
\end{figure} 

\begin{figure}
\label{Figure5}
\includegraphics[width=70mm]{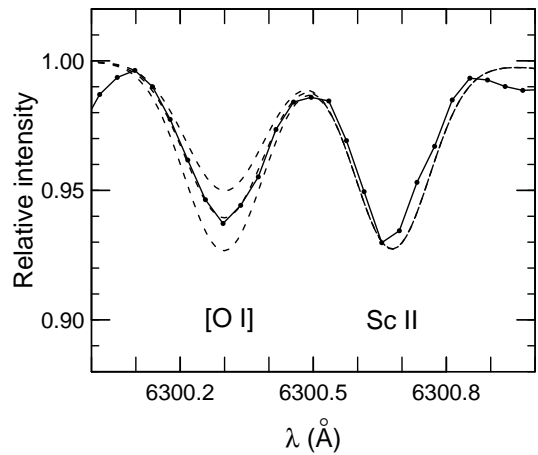}
\caption{
Synthetic spectrum fit to the forbidden [O\,{\sc i}] line at 6300~{\AA}. The observed spectra
are shown as solid line. The  dashed lines are synthetic spectra with ${\rm [O/Fe]} = 0.03$, 0.13, and 0.23~dex.
}
\end{figure} 

The abundance of Na\,{\sc i} was estimated using the line 5148.84~\AA\
which, due to rotational broadening, is blended by the Ni\,{\sc i} line at
5148.66~\AA.  These two lines are distinct in the Sun, so we were able
to calibrate their log\,$gf$ values using the solar spectrum.  However,
the sodium abundance in our study is affected by the uncertainty of
the nickel abundance, originating from the equivalent
widths method.  Fortunately, the line-to-line abundance scatter of [Ni/H]
from 24 lines of Ni\,{\sc i} was as small as 0.06~dex.

\begin{figure}
\label{Figure6}
\includegraphics[width=70mm]{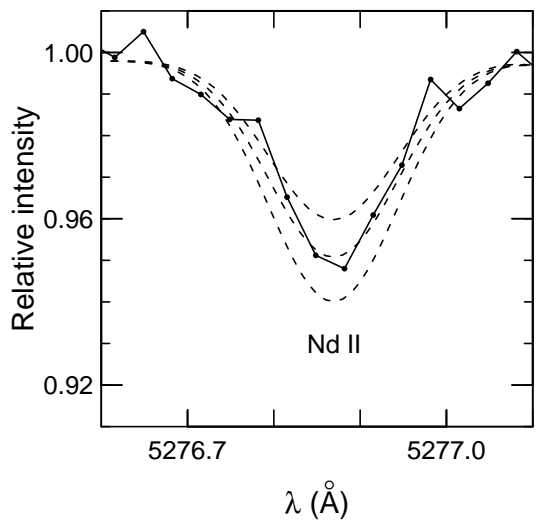}
\caption{
Synthetic  spectrum fit to the Nd\,{\sc ii} line at
5276.806~{\AA}.  The observed spectrum is shown as a solid
line. The  dashed lines are synthetic spectra  with
${\rm [Nd/Fe]} = 0.11$, 0.21, and 0.31~dex.
}
\end{figure}

\begin{figure}
\label{Figure7}
\includegraphics[width=70mm]{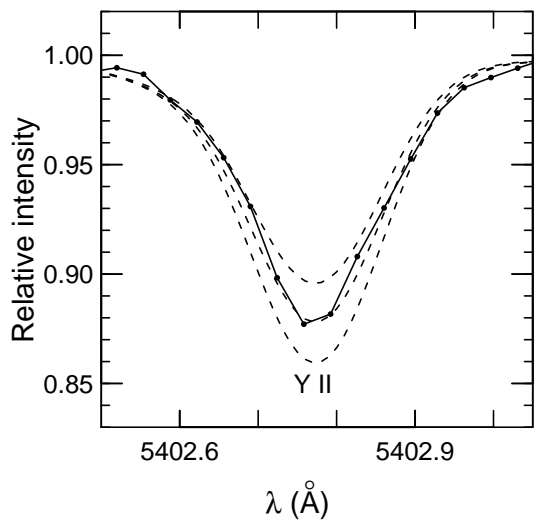}
\caption{
Synthetic  spectrum fit to the Y\,{\sc ii} line at 5402.78~{\AA}.
  The observed spectrum is shown as a solid
line. The  dashed lines are synthetic spectra  with
 ${\rm [Y/Fe]} = -0.07$, 0.03 and 0.13~dex.
}
\end{figure}

For the evaluation of the Zr\,{\sc i} abundance the lines at 5385.13~\AA,
6127.48~\AA, and 6134.57~\AA\ were used.  The abundance of Y\,{\sc ii}
was based on 5402.78~\AA, that of La\,{\sc ii} on 6390.48~\AA, that of Ce\,{\sc ii} 
on 5274.22~\AA\ and 6043.38~\AA,  and that of Nd\,{\sc ii} on 5276.86~\AA.
The synthetic spectrum fits to the Nd\,{\sc ii} line at
5276.806~\AA,  Y\,{\sc ii} at 5402.78~\AA,  and  La\,{\sc ii} at
6390.48~\AA\ are displayed in Figs.~6, 7, and 8 respectively.

The abundance of $r$-process element praseodymium was based on the  Pr\,{\sc ii} 
line at 5259.72~\AA\  and of europium on the Eu\,{\sc ii} line at 6645.10~\AA\ 
(Fig.~9).  The hyperfine structure of Eu\,{\sc ii} was taken into
account when calculating the synthetic spectrum.  The wavelength,
excitation energy and total log\,$gf = 0.12$ were taken
from Lawler et
al.\ (2001), the isotopic meteoritic fractions of $^{151}{\rm Eu}$,
47.77\,\%, and $^{153}{\rm Eu}$, 52.23\,\%, and isotopic shifts were taken
from Biehl (1976).

\subsection{Estimation of uncertainties}
The sources of uncertainty were described in detail in Paper~I.
The sensitivity of the abundance estimates to changes in the atmospheric
parameters for the assumed errors ($\pm100$~K for $T_{\rm eff}$, 
$\pm0.3$~dex for log\,$g$, and $\pm0.3~{\rm km\,s}^{-1}$ for $v_{\rm t}$) is
illustrated in Table~2.  It is seen that possible parameter errors do
not affect the abundances seriously; the element-to-iron ratios, which
we use in our discussion, are even less sensitive. The $^{12}$C/$^{13}$C ratio is particularly not
sensitive to changes in the atmospheric parameters. However, it's value might decrease by -3 or increase by +7 as a result of raising or lowering the  continuum level, respectively. The continuum level placement depends on the S/N ratio, which is about 200 in our case.

\begin{figure}
\label{Figure8}
\includegraphics[width=70mm]{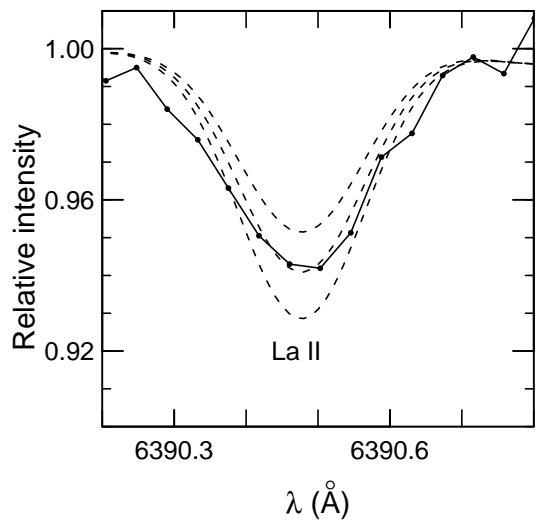}
\caption{
Synthetic  spectrum fit to the La\,{\sc ii} line at
at 6390.48~{\AA}.  The observed spectrum is shown as a solid
line. The  dashed lines are synthetic spectra  with
${\rm [La/Fe]} = -0.07$, 0.03, and 0.13~dex.
}
\end{figure}

\begin{figure}
\label{Figure10}
\includegraphics[width=70mm]{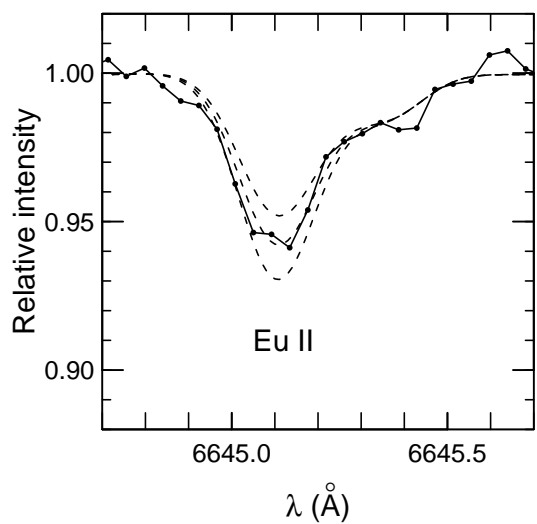}
\caption{
Synthetic spectrum fit to the Eu\,{\sc ii}  line at
6645.10~{\AA}. The observed spectrum
is shown as a solid line. The dashed lines  are synthetic spectra with
${\rm [Eu/Fe]} = 0.05$, 0.15, and 0.25~dex.
}
\end{figure}

The scatter of the deduced line abundances, $\sigma$, presented in
Table~3, gives an estimate of the uncertainty due to the random errors,
e.g., in the continuum placement and the line parameters (the mean value
of $\sigma$ is 0.05~dex).  Thus the uncertainties in the derived
abundances originating from the random errors are close to this value.

Since the abundances of C, N, and O are bound together by the molecular
equilibrium, we have also investigated how an error in one of them
typically affects the abundances  of the others.

The error $\Delta{\rm [O/H]}=0.10$ causes
$\Delta{\rm [C/H]}=0.03$ and $\Delta{\rm [N/H]}=0.05$;
$\Delta{\rm [C/H]}=0.10$ causes $\Delta{\rm [N/H]}=-0.10$ and
$\Delta{\rm [O/H]}=0.02$. However, $\Delta {\rm [N/H]}=0.10$ has no effect
on both the carbon and oxygen abundance.

\begin{table}
\caption{The sensitivity of stellar atmospherice abundances to changes of the atmospheric parameters. 
The table entries show $\Delta$\,[A/H] for the given changes of the stellar parameters.}
\label{Table2}
\tabcolsep=12pt
\begin{tabular}{lrrc}
\hline\noalign{\smallskip}
Species &$\Delta T_{\rm eff}$&$\Delta {\rm log}\rlap{\,$g$}$&$\Delta v_{\rm t}$\\[1.5pt]
 &$+100$ \rlap{K}~ & $+0.3$ & $+0.3\, {\rm km\,s}^{-1}$ \\[1.5pt]
\hline\noalign{\smallskip}
C(C$_2$)          & $0.02  $ & $0.04  $ & $0.01  $  \\
N(CN)             & $0.10  $ & $0.06  $ & $0.01  $  \\
O([O\,{\sc i}])   & $0.01  $ & $0.14  $ & $0.01  $  \\
Na\,{\sc i}       & $0.08  $ & $-0.01 $ & $\llap{$-$}0.01 $  \\
Si\,{\sc i}       & $0.01  $ & $0.05  $ & $\llap{$-$}0.03 $  \\
Ca\,{\sc i}       & $0.07  $ & $-0.01 $ & $\llap{$-$}0.04 $  \\
Sc\,{\sc ii}      & $-0.01 $ & $0.14  $ & $\llap{$-$}0.07 $  \\
Ti\,{\sc i}       & $0.12  $ & $0.00  $ & $\llap{$-$}0.02 $  \\
V\,{\sc i}        & $0.13  $ & $0.00  $ & $\llap{$-$}0.02 $  \\
Cr\,{\sc i}       & $0.07  $ & $0.00  $ & $\llap{$-$}0.03 $  \\
Fe\,{\sc i}       & $0.07  $ & $0.01  $ & $\llap{$-$}0.05 $  \\
Fe\,{\sc ii}      & $-0.06 $ & $0.16  $ & $\llap{$-$}0.05 $  \\
Co\,{\sc i}       & $0.09  $ & $0.04  $ & $\llap{$-$}0.02 $  \\
Ni\,{\sc i}       & $0.06  $ & $0.04  $ & $\llap{$-$}0.06 $  \\
Y\,{\sc ii}       & $-0.01 $ & $0.13  $ & $\llap{$-$}0.02 $  \\
Zr\,{\sc i}       & $0.16  $ & $-0.01 $ & $0.01  $  \\
La\,{\sc ii}      & $0.02  $ & $0.14  $ & $\llap{$-$}0.01 $  \\
Ce\,{\sc ii}      & $0.02  $ & $0.13  $ & $\llap{$-$}0.02 $  \\
Pr\,{\sc ii}      & $0.02  $ & $0.14  $ & $\llap{$-$}0.01 $  \\
Nd\,{\sc ii}      & $0.01  $ & $0.14  $ & $\llap{$-$}0.01 $  \\
Eu\,{\sc ii}      & $-0.01 $ & $0.14  $ & $0.01  $  \\
\\[-6pt]
C/N &	$	-1.73	$ & $	-0.49	$ & $	0.18 $ \\
\textsuperscript{12}C/\textsuperscript{13}C & $	-2 $ & $ 1 $ & $ -2 $  \\[1.5pt]
\hline
\end{tabular}
\end{table}

\section{Results and discussion}

From the observed spectrum we determined for AY\,Cet the following main atmospheric
parameters:  
${T_{\rm eff}=5080}$~K, 
$\log g=3.0$, $v_{\rm t}=1.4~{\rm km}\,s^{-1}$, 
and ${\rm [Fe/H]}=-0.33$. 

Similar values of effective temperatures for AY\, Cet (lying in the 
interval from 5000~K to 5100~K) and of log\,$g$ (from 2.8 till 3.1)  
have been determined by Berdyugina (1994), Biazzo et al.\ (2007), da Silva et al.\ (2006), 
Pallavicini et al.\ (1992), and Randich et al.\ (1993).  Identical ${\rm [Fe/H]}=-0.33$ for AY\, Cet 
have been determined also in the recent studies of  Biazzo et al.\ (2007) and da Silva et al.\ (2006). 
Lower values of the metallicity were determined in the earlier studies of Randich et al.\ (--0.5~dex, 1993) 
and  Berdyugina (--0.58~dex, 1994). 
The abundances [A/H], C/N, and \textsuperscript{12}C/\textsuperscript{13}C together with their
$\sigma$ (the line-to-line scatter) are listed in Table~3. 

\begin{table}
\caption{Atmospheric abundances relative to hydrogen, [A/H]. $\sigma$ is the standard deviation of
 the mean value due to the line-to-line scatter; 
 $N$ is the number of lines used for the abundance determination.}
\label{Table3}
\tabcolsep=15pt
\begin{tabular}{lrrc}
\hline\noalign{\smallskip}
Species & $N$ & [A/H] & $\sigma$ \\[1.5pt]
\hline\noalign{\smallskip}
C(C$_2$)        & 1   & $-0.41 $ & $-     $  \\
N(CN)           & 4   & $-0.01 $ & $0.05  $  \\
O([O\,{\sc i}]) & 1   & $-0.17 $ & $-     $  \\
Na\,{\sc i}     & 1   & $-0.16 $ & $-     $  \\
Si\,{\sc i}     & 7   & $-0.29 $ & $0.03  $  \\
Ca\,{\sc i}     & 4   & $-0.22 $ & $0.06  $  \\
Sc\,{\sc ii}    & 5   & $-0.22 $ & $0.06  $  \\
Ti\,{\sc i}     & 18  & $-0.17 $ & $0.04  $  \\
V\,{\sc i}      & 19  & $-0.15 $ & $0.05  $  \\
Cr\,{\sc i}     & 7   & $-0.23 $ & $0.05  $  \\
Fe\,{\sc i}     & 49  & $-0.33 $ & $0.04  $ \\
Fe\,{\sc ii}    & 6   & $-0.33 $ & $0.06  $ \\
Co\,{\sc i}     & 8   & $-0.37 $ & $0.07  $ \\
Ni\,{\sc i}     & 24  & $-0.42 $ & $0.06  $ \\
Y\,{\sc ii}     & 1   & $-0.23 $ & $-     $ \\
Zr\,{\sc i}     & 2   & $-0.17 $ & $0.01  $ \\
La\,{\sc ii}    & 1   & $-0.15 $ & $-     $ \\
Ce\,{\sc ii}    & 2   & $-0.29 $ & $0.04  $ \\
Pr\,{\sc ii}    & 1   & $-0.25 $ & $-     $ \\
Nd\,{\sc ii}    & 1   & $-0.03 $ & $-     $ \\
Eu\,{\sc ii}    & 1   & $-0.18 $ & $-     $ \\
\\[-6pt]
C/N              & 5 & $1.58$ & 0.17 \\ 
\textsuperscript{12}C/\textsuperscript{13}C & 1 & 21 & $-$\\[1.5pt]
\hline
\end{tabular}
\end{table}
\begin{figure*}
\label{Figure10}
\includegraphics[width=150mm]{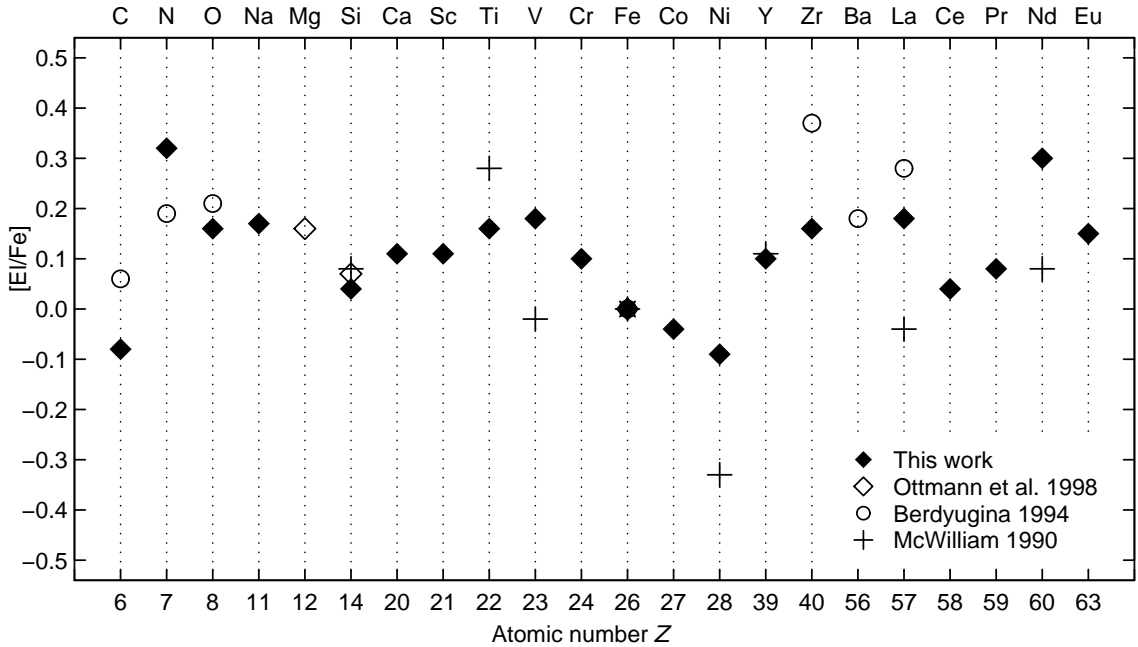}
\caption{
Element abundances for AY\,Cet, as determined in this work
(filled diamonds), by Ottmann et al.\ (1998) (open diamonds), by Berdyugina (1994) (circles), and by McWilliam (1990) (plus signs).
}  
\end{figure*}

\begin{figure*}
\label{Figure11}
\includegraphics[width=158mm]{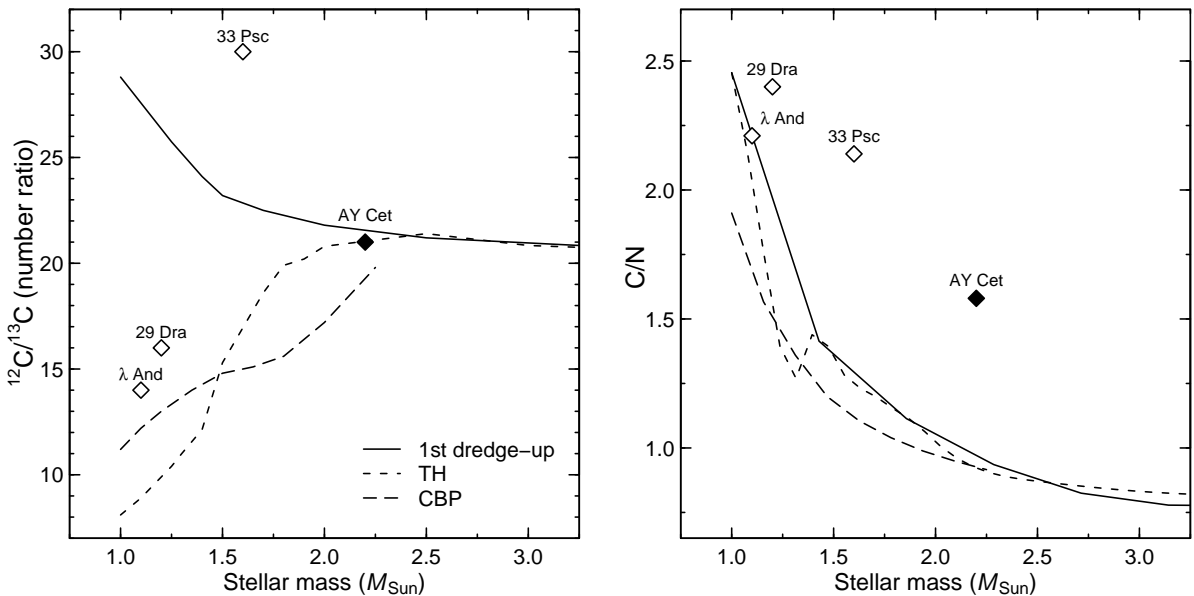}
\caption{
Comparisons of $^{12}$C/$^{13}$C and C/N ratios in RS~CVn stars with theoretical models. 
AY\,Cet  is marked by the filled diamond, other RS~CVn stars by open diamonds: 
33\,Psc (Barisevi\v{c}ius et al.\ 2011), $\lambda$\,And (Paper~I), and 29\,Dra (Barisevi\v{c}ius et al.\ 2010).  
The theoretical models are explained in the text.
}  
\end{figure*}

The values of [El/Fe] obtained for AY\,Cet are displayed 
in Fig.~10 together with results of other analyses of this star. 
Abundances of C, N, O, Li, Zr, Ba, and La in AY\,Cet were also previously determined by Berdyugina (1994), 
and Mg and Si  were investigated by Ottmann et al.\ (1998). These results are in quite  good agreement with ours. 
The elements Si, Ti, V, Ni, Y, La, and Nd  were investigated by McWilliam (1990). The abundances of Si, Ti, and Y are close to ours, but
the abundances of the other elements are  about 0.3~dex lower. Very different results (not displayed in Fig.~10) 
were obtained 
by Sanz-Forcada et al.\  (2009). Their [O/Fe] value  is equal to 1.25~dex, and also the other element-to-iron 
ratios are as high as 0.4 or 0.8~dex. The log\,$g$ value in their study is lower by about 0.7~dex than in 
other studies, the effective temperature is higher by about 200~K, and [Fe/H] = 0.0. The spectrum of AY\,Cet 
for their analysis was observed in November of 2002. 
 
The evolutionary sequences in the 
Hertzsprung-Russell-diagram of Girardi et al.\ (2000) show that AY\,Cet (with 
 ${{\rm log}(L/L_{\odot})=1.65}$ and  $\sim\!2.2$\,M$_{\odot}$) is a first ascent giant lying
close to the bottom of the giant branch in the area where  the first dredge-up occurs. 
In such stars, the 1st dredge-up 
starts when the luminosity ${\rm log}(L/L_{\odot})$ is ${\approx} 1.2$ and ends at ${\approx\! 1.8}$ (Boothroyd \& Sackmann 1999). 
So, the convective envelope should not reach the deepest layers of AY\,Cet yet.

In Fig.~11, we compare the C/N and $^{12}$C/$^{13}$C ratios of AY\,Cet 
with the standard prediction of the 1st dredge-up and with two models of extra-mixing.  The model called `cool
bottom processing' (CBP) was proposed by Boothroyd \& Sackmann (1999), and the model called 'thermohaline mixing' (TH) 
was proposed by Charbonnel \& Lagarde (2010).

The position of AY\,Cet in the C/N versus
stellar mass diagram (Fig.~11) shows a higher C/N ratio as compared to the 1st dredge-up prediction, 
as expected from the evolutionary status of AY\,Cet. 
Concerning the  $^{12}$C/$^{13}$C ratio of AY\,Cet, 
it is as low as at the end of the 1st dredge-up. 

The abundance of lithium is also very sensitive to mixing.
During the first dredge-up, for a star of the mass of AY\,Cet, the Li abundance
drops to approximately log\,$A{\rm (Li)}=1.36$ (Charbonnel \& Lagarde
2010).  However, the available determinations of log\,$A{\rm (Li)}$  in AY\,Cet
show values spanning from 0 to 0.6~dex (Randich et al.\ 1993; Barrado y Navascu\'{e}s et al.\ 1998; Berdyugina 1994; 
Costa et al.\ 2002; Fekel \& Balachandran 1993).  
The determined abundances of lithium in AY\,Cet are lower than predicted by 
the first dredge-up model while the carbon and nitrogen abundances are in agreement 
with it. 
Thus any reasons of the low lithium abundance should be related to other mechanisms. 

\subsection{Comparison of photospheric and coronal abundances}

One of the still unanswered questions in stellar astrophysics is weather the coronal abundances
 in cool stars are 
similar to the photospheric ones. In the Sun, an enhancement of elements 
with a low first ionization potential (FIP) is found in the corona with respect to the photosphere. 
The different composition in the corona and the photosphere may indicate that some physical 
processes take place between the cooler photospheric material and the hotter corona. Sanz-Forcada et al.\ (2009) 
have presented a comparison of photospheric and coronal abundances in 12 stars. It is seen that some stars 
have the FIP effect, some no and some have even an inverse FIP effect. 

Comparing the photospheric abundances of AY\,Cet as determined in this work with the coronal abundances 
(Sanz-Forcada et al.\ 2009), we see no FIP effect. It is interesting to notice that the abundances of O and Si 
are very close, the abundance of C in the corona is higher by about 0.8~dex, of Ni by 0.5~dex,  of N by 0.4~dex, 
and that of Fe is lower by 0.15~dex. More observational data are certainly needed in order to reveal the true 
 abundance differences in stellar photospheres and coronae.  


\begin{thebibliography}{}
\bibitem{}  Alonso,~A., Arribas,~S., Mart\'{i}nez-Roger,~C.: 1999, A\&AS 140, 261
\bibitem{} Barisevi\v{c}ius,~G., Tautvai\v{s}ien\.{e},~G., Berdyugina,~S., Chorniy,~Y., Ilyin,~I.: 2010, Baltic Astronomy 19, 157
\bibitem{} Barisevi\v{c}ius,~G., Tautvai\v{s}ien\.{e},~G., Berdyugina,~S., Chorniy,~Y., Ilyin,~I.: 2011, Baltic Astronomy 20, 53
\bibitem{} Barrado y Navascu\'{e}s, D., De Castro, E., Fern\'{a}ndez-Figueroa, M.J., Cornide, M., 
Garci\'{a},~R.~J.: 1998, A\&A 337, 739
\bibitem{} Batten,~A.H., Fletcher,~J.M., MacCarthy,~D.G.: 1989, Publications of the Dominion Astrophysical Observatory 17, 1
\bibitem{} Berdyugina,~S.V.: 1994, Astronomy Letters 20, 796
\bibitem{} Biazzo,~K., Pasquini,~L., Girardi,~L., et al.: 
2007, A\&A 475, 981
\bibitem{} Biehl,~D.: 1976, Diplomarbeit, Christian-Albrechts-Universit\"at Kiel, Institut f\"ur 
Theoretische Physik und Sternwarte
\bibitem{} Boothroyd,~A.I., Sackman,~I.J.: 1999, ApJ 510, 232
\bibitem{} Charbonnel,~C., Lagarde,~N.: 2010, A\&A 522, A10
\bibitem{} Costa,~J.M., da~Silva,~L., do~Nascimento~Jr.,~J.D., De~Medeiros, J.R.: 2002, A\&A 382, 1016
\bibitem{} Cousins,~A.W.J.: 1962, Mon. Notes Astron. Soc. S. Afr. 21, 20
\bibitem{} Cowley,~A.P.,  Bidelman,~W.P.: 1979, PASP 91, 83
\bibitem{} da Silva,~L., Girardi,~L., Pasquini,~L., et al.:  
2006, A\&A 458, 609
\bibitem{} Eaton,~J.A., Hall,~D.S., Henry, G.W., et al.:  
1983, Ap\&SS 93, 271
\bibitem{} Fekel,~F.C., Eitter,~J.J.: 1989, AJ 97, 1139 
\bibitem{} Fekel,~F.C., Balachandran,~S.: 1993, AJ 403, 708
\bibitem{} Fekel,~F.C., Moffett,~T.J., Henry,~G.W.: 1986, ApJS 60, 551

\newpage
\bibitem{} Girardi,~L., Bressan,~A., Bertelli,~G., Chiosi,~C.: 2000, A\&AS 141, 371
\bibitem{} Gray,~D.F.: 1989, ApJ 347, 1021
\bibitem{} Grevesse,~N., Sauval,~A.J.: 2000, in: O. Manuel (ed.), {\it Origin of Elements in the Solar System, 
Implications of Post-1957 Observations},  p.\,261

\bibitem{} Hakkila,~J., Myers,~J.M., Stidham,~B.J., Hartmann,~D.H.: 1997, AJ 114, 2043
\bibitem{} Hauck,~B., Mermilliod,~M.: 1998, A\&AS 129, 431
\bibitem{} Johansson,~S., Litzen,~U., Lundberg,~H., Zhang,~Z.: 2003, ApJ 584, 107
\bibitem{} Kurucz,~R.~L.: 2005, 
Mem. Soc. Astron. Ital. Suppl. 8, 189
\bibitem{} Lawler,~J.E., Wickliffe,~M.E., Den Hartog,~E.A.: 2001, ApJ 563, 1075
\bibitem{} Massarotti,~A., Latham,~D.W., Stefanik,~R.P., Fogel,~J.: 2008, AJ 135, 209
\bibitem{} McWilliam,~A.: 1990, ApJS 74, 1075
\bibitem{} Mermilliod,~J.C.: 1986, Catalogue of Eggen's UBV data
\bibitem{} Montes,~D., Fernandez-Figueroa,~M.J., de~Castro,~E., Cornide,~M.: 1994, A\&A 285, 609
\bibitem{} Olsen,~E.H.: 1974, IBVS 925
\bibitem{} Ottmann,~R., Pfeiffer,~M.J., Gehren,~T.: 1998, A\&A 338, 661
\bibitem{} Pallavicini,~R., Randich,~S., Giampapa,~M.S.: 1992, A\&A 253, 185
\bibitem{} Poretti,~E., Mantegazza,~L., Hall,~D.S. et al.: 1986, A\&A 157, 1
\bibitem{} Randich,~S., Gratton,~R., Pallavicini,~R.: 1993, A\&A 273, 194
\bibitem{} Sanz-Forscada,~J., Affer,~L., Micela,~G.: 2009, A\&A 505, 299
\bibitem{} Schrijver,~C.J., Zwaan,~C.: 1991, A\&A 251, 183
\bibitem{} Simon,~T., Fekel,~F.C.~Jr., Gibson,~D.M.: 1985, ApJ 295, 153
\bibitem{} Simon,~T., Sonnerborn,~G.: 1987, AJ 94, 1657 
\bibitem{} Strassmeier,~K.G.,  Hall,~D.S., Zeilik,~M., Nelson,~E., Eker,~Z., Fekel,~F.C.: 1988, A\&AS 72, 291
\bibitem{} Strassmeier,~K.G., Hall,~D.S., Boyd,~L.J., Gent,~M.: 1989, ApJS 69, 141
\bibitem{} Tautvai\v{s}ien\.{e},~G., Barisevi\v{c}ius,~G., Berdyugina,~S., Chorniy,~Y., 
Ilyin,~I.: 2010, Baltic Astronomy 19, 95 (Paper~I)
\bibitem{} van Leeuwen,~F.: 2007, {\it Hipparcos, the New Reduction of the Raw Data}, 
Astrophysics and Space Science Library 350
\bibitem{} Walter,~F.M.,  Bowyer,~S.: 1981, ApJ 245, 671

\end{thebibliography}
\end{document}